# Enhanced anomalous Hall effect in the topological Kagome metal Cs(V$_{1-x}$Mn$_x$)$_3$Sb$_5$

Xinmin Wang[1,2,3], Peipei Wang[4,5], Jian Lyu[1,2,6], Zhuang Xu[1,2,3], Mingquan He[7], Yu Feng[1,2,3], Wei Luo[1,2,3], Liyuan Zhang[4,5], Junying Shen[1,2,3,*], and Xin Tong[1,2,3,*].

[1] *Spallation Neutron Source Science Center, Dongguan 523803, China*

[2] *Institute of High Energy Physics, Chinese Academy of Sciences (CAS), Beijing 100049, China*

[3] *Guangdong Provincial Key Laboratory of Extreme Conditions, Dongguan 523803, China*

[4] *Department of Physics of Southern University of Science and Technology, Shenzhen 518055, China*

[5] *Quantum Science Center of Guangdong Hong Kong Macao Greater Bay Area (Guangdong), Shenzhen 51805, China*

[6] *Research and Development Center for Ecological Environment Engineering Technology, Dongguan University of Technology, Dongguan 523808, China*

[7] *College of Physics & Center of Quantum Materials and Devices, Chongqing University, Chongqing 401331, China*

## Abstract

As a fundamental physical phenomenon, achieving and controlling a large anomalous Hall effect (AHE) is crucial for advancing the understanding of topological physics and for developing applied technologies in spintronics. The recently discovered topological Kagome metal $A$V$_3$Sb$_5$ ($A$ = K, Rb, Cs) exhibits a significant AHE along with charge density wave (CDW) and superconductivity, providing an ideal platform to study the interactions between nontrivial band topology, CDW, and superconductivity. In this study, we systematically investigated the evolution of CDW, superconductivity, and AHE in electron (Mn)-doped Cs(V$_{1-x}$Mn$_x$)$_3$Sb$_5$ single crystals. The experimental results show that electron doping rapidly suppresses superconductivity, while the CDW order remains relatively robust. Meanwhile, a significantly enhanced AHE, with a maximum anomalous Hall conductivity (AHC) of ~25331 $\Omega^{-1}$cm$^{-1}$ and an anomalous Hall angle of ~6.66% occurs at a relatively low doping level of x = 0.03. Based on the Tian–Ye–Jin (TYJ) scaling model, such a significant enhancement AHC is mainly dominated by the skew scattering. We speculated enhanced skew scattering between electrons and Mn originating from the strengthened spin-orbital coupling. Our finding provides important guidance for the design and development of transverse transport properties in topological Kagome materials.

## Introduction

The recently discovered Kagome metal $A$V$_3$Sb$_5$ ($A$ = K, Rb, Cs) has attracted much attention due to its complex electronic band structure primarily manifesting as flat bands, van Hove singularities (vHs), and Dirac points[1-3]. These novel structures lead to many exotic quantum phenomena in $A$V$_3$Sb$_5$, such as unconventional superconductivity[4, 5], chiral charge density waves (CDWs)[6, 7], anomalous Hall effects (AHE)[8-10], electronic nematic phases[11-13], and pair density waves[14] and so on. Among these, the giant AHE holds particular significance because it reveals key insights into the topological properties and magnetic interactions. Multiple mechanisms have been proposed to explain the unconventional giant AHE in $AV_3Sb_5$, including large Berry curvature[10], the enhanced skew scattering of the CDW state or "spin cluster"[8, 9], the orbital currents of novel chiral charge order[7], et. Moreover, the AHE can be tuned through various ways including pressure[15-18], dimension[19], gating[10, 20, 21], and chemical doping[22-25], providing valuable information about the underlying mechanisms. These methods can effectively engineer the electronic band structure, thereby altering the intrinsic contribution to the AHE by directly modifying the Berry curvature strength around the Fermi surface, as observed in Ni-doped $Co_{3-x}Ni_xSn_2S_2$[26], gated-$MnBi_2Te_4$ films[20] and pressured CeAlSi[17]. In addition, the introduction of alien atoms would enhance the scattering of conduction electrons, resulting in the change of extrinsic anomalous Hall conductivity (AHC) in the system[27, 28]. As in $Co_{3-x}Fe_xSn_2S_2$, the asymmetric scattering of moving electrons due to Fe dopants contributes to the extrinsic effect, leading to a sizable enhancement of AHC[28]. Therefore, by utilizing suitable impurities, both extrinsic and intrinsic contributions can be leveraged to enhance the desired AHC.

There are several studies on the doping effect of $CsV_3Sb_5$, mainly focusing on the influence of doping on CDW and superconductivity, such as $Cs(V_{1-x}B_x)_3Sb_5$ (B = Ti, Cr, Nb, Mo, Ta) and $CsV_3Sb_{5-x}Sn_x$[29-33]. However, a detailed study on the transverse Hall effect of doping $CsV_3Sb_5$ is rare. It is reported that the AHE is monotonically suppressed in $Cs(V_{1-x}Cr_x)_3Sb_5$[31] and $Cs(V_{1-x}Mo_x)_3Sb_5$[34]. Interestingly, an extrinsic-intrinsic transition of AHE was achieved in $CsV_3Sb_5$ nanodevices though electrically controlled proton intercalation[10]. Here, we report an enhanced AHE in Mn-doped $Cs(V_{1-x}Mn_x)_3Sb_5$ and study the evolution of superconductivity and CDW order. Electrical transport and magnetization measurements indicate that partial substitution of V with Mn simultaneously suppresses both superconductivity and CDW order. Hall measurements reveled that a small among of doped Mn atoms randomly occupy V sites in the Kagome lattice leading to a sizable enhancement of AHC of up to 20% compared with the undoped sample. Scaling model shows that such an enhancement of AHC is mainly from enhanced extrinsic skew scattering due to the strengthened of spin-orbital coupling (SOC) of conduction electrons and Mn impurities, and the intrinsic contributions almost unchanged. This finding provides an effective method to enhance the AHE in topological Kagome metals.

## Experiments and Methods

Single crystals of $Cs(V_{1-x}Mn_x)_3Sb_5$ with a range of Mn-doping concentrations were prepared by the self-flux method. The starting materials with a molar ratio of 1:3:20 were put into an alumina crucible and vacuum-sealed in a quartz tube. This tube was gradually heated to 1000°C and maintained for 20 hours, then cooled to 850°C at a rate of 100°C/h, and further to 680°C at 1.5°C/h. Subsequently, centrifugation was employed to separate the crystals from the flux, yielding hexagonal plate-shaped single crystals with minimal air sensitivity. Owing to the high reactivity of the initial materials, all preparation steps were performed under an inert argon gas atmosphere.

The crystal structure was verified using X-ray diffraction (XRD) on a PANalytical diffractometer with Cu Kα radiation at room temperature. The magnetic properties were measured by Quantum Design Magnetic Properties Measurement System. The electrical transport characteristics were respectively quantified using a Physical Properties Measurement System. All samples used for transport measurements were mechanically exfoliated to remove the oxidized layer and then carefully cut into regular sticks.

# Results and Discussion

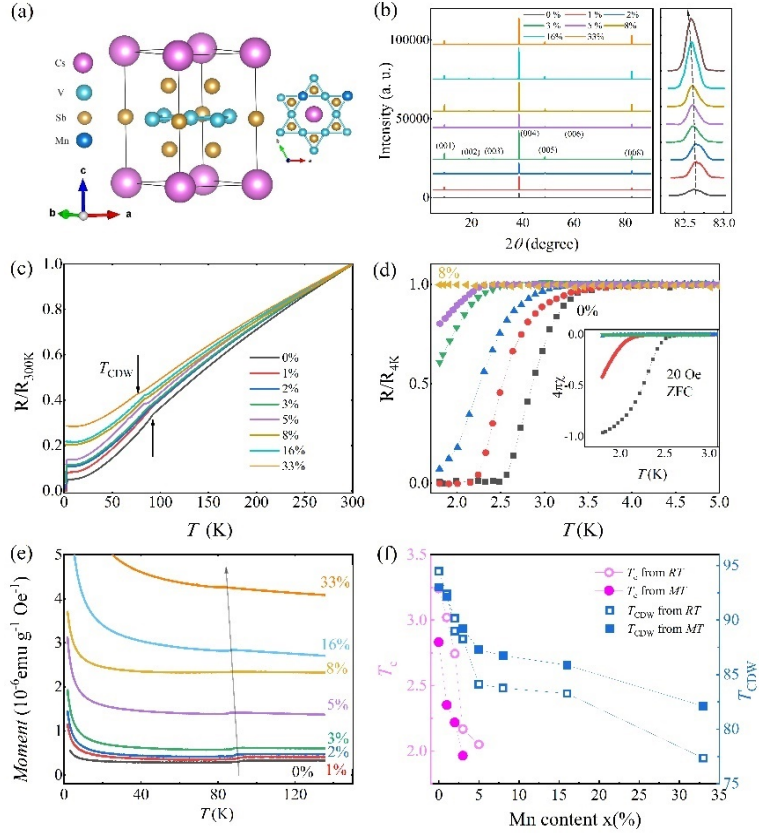

FIG. 1. (a) The crystal structure of Mn-doped $Cs(V_{1-x}Mn_x)_3Sb_5$ and a Kagome layer composed of V/Mn atoms. (b) The XRD patterns for $Cs(V_{1-x}Mn_x)_3Sb_5$. (c) - (d) The temperature dependence of $\rho_{xx}$ for different Mn contents. The data were normalized with respect to the resistivity at 300 K. The arrows indicate the CDW transition. The inset of (d) The temperature dependence of magnetic susceptibility was measured with an applied field of 20 Oe in the *ab* plane to show the evolution of $T_c$. (e) Temperature dependence of magnetic susceptibility for different Mn content. The arrows indicate the CDW transition. (f) The phase diagram of $Cs(V_{1-x}Mn_x)_3Sb_5$ with $T_c$ and $T_{CDW}$ from *R*-T and *M*-T results.

The crystal structure of $Cs(V_{1-x}Mn_x)_3Sb_5$, as shown in Figure 1(a), consists of a quasi-two-dimensional Kagome lattice formed by V and Mn. X-ray single crystal diffraction patterns of $Cs(V_{1-x}Mn_x)_3Sb_5$ are presented in Figure 1(b). All peaks can be identified as the (00*l*) reflections of $CsV_3Sb_5$ as labeled on the pattern, indicating that Mn-doping does not affect the crystal structure, similar to the phenomenon in $Cs(V_{1-x}Cr_x)_3Sb_5$[31]. For clarity, the diffraction peaks of (008) were amplified, and it was found that as the Mn doping concentration increased, the peak slightly shifted left, indicating that Mn was successfully substituted for V-site in the Kagome layer. Electronic transport and magnetic properties measurements were performed to characterize the interaction between superconductivity and CDW order. Figure 1(c) shows the temperature dependence of resistivity of $Cs(V_{1-x}Mn_x)_3Sb_5$ with x = 0, 0.01, 0.02, 0.03, 0.05, 0.08, 0.16, 0.33, respectively. As the Mn increases, CDW transition temperature monotonically decreases from 93.6 K for parent $CsV_3Sb_5$ to 78 K for the sample with x = 0.33. With further increasing Mn dopants, the CDW transition temperature is almost unchanged, indicating the CDW order is robust against electronic doping. A similar shift associated with the CDW transition is also detected in the temperature dependence of magnetic susceptibility measured under zero field cooling (ZFC) model with an applied field $\mu_0H$ =1 T parallel to ab plane for all Mn-doped Cs(V$_{1-x}$Mn$_x$)$_3$Sb$_5$ samples, as shown in Fig. 1(e). Furthermore, the superconducting transition was significantly suppressed as the Mn doping content x increased, as indicated in Fig. 1(d). We also summarized the superconductivity transition temperature, $T_c$ and CDW transition temperature, $T_{CDW}$ as the function of doping level x in Fig. 1(f), here $T_c$ is defined as the temperature corresponding to a 10% decrease in resistivity and magnetic susceptibility. It is clearly shown that the introduction of Mn can kill the superconductivity very quickly and slightly weaken CDW order, revealing an unconventional coexisting mechanism between superconductivity and CDW, similar to Cs(V$_{1-x}$Cr$_x$)$_3$Sb$_5$[31].

To further elucidate the properties of carriers and charge transport mechanism, we have performed the Hall resistivity measurements at various temperatures for Mn-doped Cs(V$_{1-x}$Mn$_x$)$_3$Sb$_5$, and the results are shown in Figure 2. $\rho_{xy}$ for all samples exhibit a linear behavior at high temperatures, indicating that the carrier density is dominated by one band. As the temperature further decreases, $\rho_{xy}$ slightly deviates from the linear behavior, and the negative slope of $\rho_{xy}$ changes to positive, indicating that the charge carriers have changed from electronic to hole type, which is consistent with the previous reports[9, 34]. A single band model has been employed to analyze the Hall data. The carrier concentration is given by $n = 1/eR_H$, and the Hall coefficient $R_H$ is obtained by $R_H = \rho_{xy}(9T)/H$. Figure 2(h) shows the carrier concentration vs temperatures for all Cs(V$_{1-x}$Mn$_x$)$_3$Sb$_5$ samples.

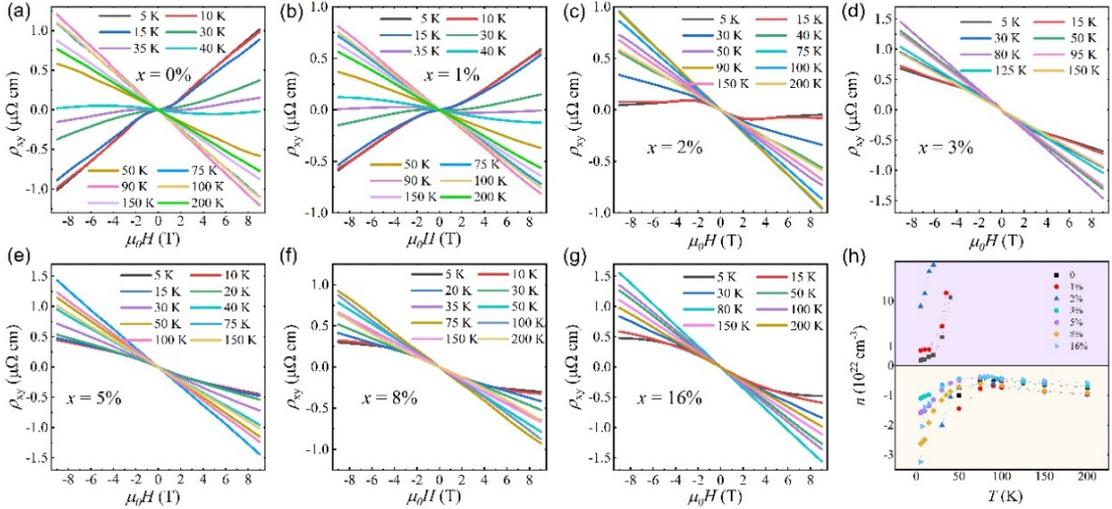

FIG. 2. (a)–(g) The field dependence of Hall resistivity at various temperatures for different Mn contents from 5 K to 200 K. (h) The temperature dependence of carrier density for different Mn contents.

It is found that the carrier density is smallest at $T_{CDW}$ for all samples, which is agreement with earlier reports[8]. As the Mn increases, the carrier-type transition temperature gradually decreases from 40 K for the parent to 20 K for the sample with x=0.02, and then the carriers are completely electron dominated during totally measurement temperature range for the sample with x≥0.03. Notably, among all doped Cs (V$_{1-x}$Mn$_x$)$_3$Sb$_5$ samples, the smallest carrier appears at x = 0.03. All results suggest that the Mn successfully substituted for V density and effectively affected the distribution of electronic states.

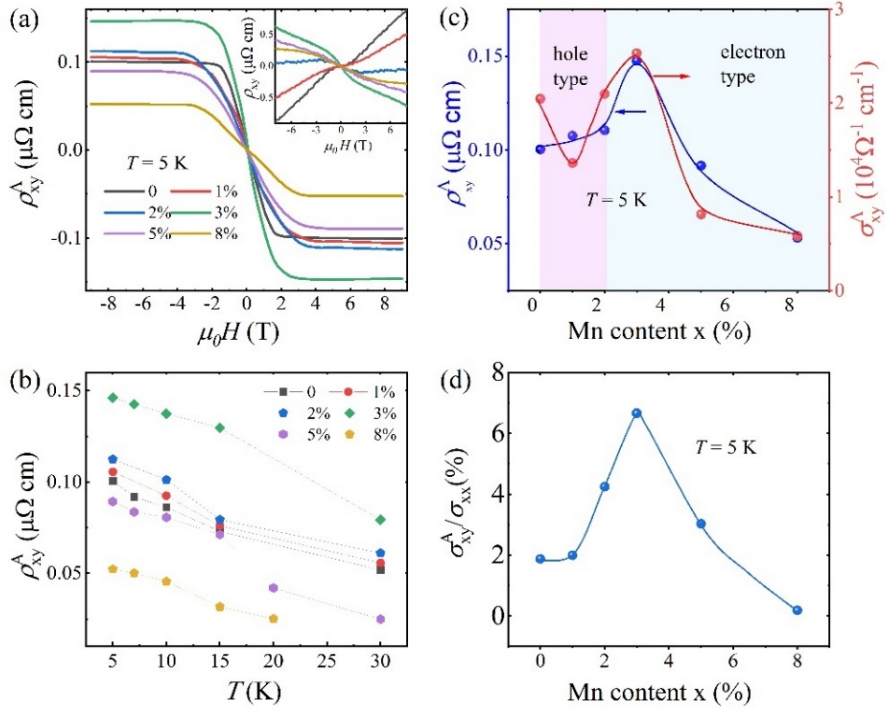

FIG. 3. (a) The extracted anomalous Hall resistivity $\rho_{xy}^A$ by subtracting the local linear ordinary Hall background at 5 K for different Mn contents. The inset shows the $\rho_{xy}$ for different Mn contents at 5 K. (b) Mn content dependence of $\rho_{xy}^A$ and $\sigma_{xy}^A$ at 5 K. (c) The temperature dependence of $\rho_{xy}$ for different Mn contents. (d) Mn content dependence of AHA at 5 K.

As shown in Figs. 2, the Hall resistivity $\rho_{xy}$ exhibits an antisymmetric sideways "S" line shape at low magnetic fields, indicating the presence of the AHE. The resistivity corresponding to the anomalous Hall effect ($\rho_{AHE}$) is obtained by subtracting the local linear ordinary Hall term, as shown in Fig. 3(a). Furthermore, we noticed that $\rho_{xy}^A$ for all doped Cs(V$_{1-x}$Mn$_x$)$_3$Sb$_5$ samples decreases with increasing temperature, as displayed in Fig. 3(b). Interestingly, $\rho_{xy}^A$ has a strong doping dependence, and we show in Fig. 3(c) the dome-like shape of $\rho_{xy}^A$ with increasing Mn content $x$ at 5 K. $\rho_{xy}^A$ reaches a maximum of 0.147 μΩ cm at $x = 0.03$, where the carrier concentration is the smallest. Then, $\rho_{xy}^A$ fades away completely with a higher doping content at $x = 0.16$, where the CDW order remains strong. According to the formula $\sigma_{xy} = -\rho_{xy}/(\rho_{xx}^2 + \rho_{xy}^2)$, the anomalous Hall conductivity (AHC) $\sigma_{xy}^A$ can be obtained and its doping dependence is also plotted together with $\rho_{xy}^A$. As the Mn content increases, $\sigma_{xy}^A$ at 5K shows a clear increase from 20,466 Ω$^{-1}$ cm$^{-1}$ at $x = 0$ to 25,331 Ω$^{-1}$ cm$^{-1}$ at $x = 0.03$, reflecting a 25% increase at low doping levels. In addition, the anomalous Hall angle (AHA) characterized by $\sigma_{xy}^A/\sigma_{xx}$ is also typically induced to study the Hall transport properties, from which the efficiency of the conversion from longitudinal current into transverse current is known. As illustrated in Fig. 3(d), the AHA appears at 5 K also shows an initial increase and a subsequent decrease, forming a maximal value of 6.66% at $x = 0.03$, which is higher than that of undoped CsV$_3$Sb$_5$.

To further quantitatively analyze the intrinsic and extrinsic contributions of AHC in Cs(V$_{1-x}$Mn$_x$)$_3$Sb$_5$, we fit the data by the so-called Tian-Ye-Jin(TYJ) model[8, 35]: $\sigma_{xy}^A = -\alpha \sigma_{xx0}^{-1} \sigma_{xx}^2 - b$, where α is the skew constant, $\sigma_{xx0} = 1/\rho_{xx0}$ is the residual conductivity, and $b$ is the intrinsic AHC contribution. We have plotted $\sigma_{xy}^A$ against $\sigma_{xx}^2$ for all Cs(V$_{1-x}$Mn$_x$)$_3$Sb$_5$ samples in Fig. 4(a). It can be clearly seen that $\sigma_{xy}^A$ exhibits a quadratic scaling relationship with $\sigma_{xx}$, which is consistent with the previous results in KV$_3$Sb$_5$[8] and CsV$_3$Sb$_5$[9]. Based on the TYJ model, the extrinsic and intrinsic contributions of AHC at 5 K were quantitatively separated, where $\sigma_{xx0} \approx \sigma_{xx}$(5 K) as shown in Fig. 4(b). It can be observed that with the increase in Mn content, the intrinsic $\sigma_{xy}^A(int.)$ increases from an initial negative contribution (-405 Ω$^{-1}$ cm$^{-1}$) to a positive contribution (1704 Ω$^{-1}$ cm$^{-1}$ at $x =$ 0.03), and then decreases with further increasing Mn content. Meanwhile, the obtained extrinsic $\sigma_{xy}^A(ext.)$ is comparable to the total AHC in magnitude with the same evolution trend. This indicates that the enhancement of AHC by small amounts of Mn doping is dominated by the extrinsic mechanisms, meanwhile the contribution from intrinsic mechanism should also be considered.

Next, we will discuss the origin of nonmonotonic change in the AHC in details. As reported in pristine CsV$_3$Sb$_5$, there are four vHs's at the M point close to the Fermi level, and Dirac bands consist of $d_{xz}/d_{yz}$ orbitals of V atoms are also closeby. It is notable that chiral CDW opens a topological energy gap near vHs and splits the bands into two sub-bands. Therefore, the Fermi level lies in CDW gap and non-zero Berry curvature, which gives rise to an anomalous Hall effect. In our case, by using Mn-doped CsV$_3$Sb$_5$, the Fermi level will be shifted upward and approach to upper sub-band, which generate a relatively large intrinsic AHE. However, by further increasing the Mn content, the Fermi level completely lies in the electron pockets and the AHE decreased gradually, as observed in Co$_{3-x}$Ni$_x$Sn$_2$S$_2$ and Co$_3$Sn$_{2-x}$In$_x$S$_2$.

As mentioned above, the significant enhancement is primarily dominated by the extrinsic skew scattering mechanism. It is worth noting that as the Mn content increases, the skew constant α rises from 0.0185 in pristine CsV$_3$Sb$_5$ to 0.071 for x = 0.03, which corresponds to the evolution of the AHC and AHA. This larger skew constant indicates that the enhanced AHC and AHA in Cs(V$_{1-x}$Mn$_x$)$_3$Sb$_5$ are mainly caused by the enhanced skew scattering. In comparison to K$_{1-x}$V$_3$Sb$_5$, which has a similar skew constant (α ~0.017), CsV$_3$Sb$_5$ shows a significant increase in AHC, rising from 15,507 Ω$^{-1}$ cm$^{-1}$ to 20,446 Ω$^{-1}$ cm$^{-1}$, an increase of more than 30%[8, 36]. This suggests that the strength of spin-orbit coupling plays a crucial role in the AHE in the AV$_3$Sb$_5$ system. Thus, we speculated that the introduction of a small amount of Mn atoms, which randomly occupy V sites in the Kagome lattice, introduces additional conduction electrons and strengthens the spin-orbit coupling between the electrons and the Mn atoms, leading to an enhancement of the AHE, as observed in Co$_{3-x}$Fe$_x$Sn$_2$S$_2$[28]. With further increase in the Mn content, the conductivity at 5 K decreases rapidly and the Kondo effect appears at x = 0.05, which is characterized by the upwarps of the $R$–T curves at low temperatures, as shown in Fig. 1(c). As shown in Fig. 4(c), the $\sigma_{xy}^A$ at x = 0.08 decreases rapidly with $\sigma_{xx}$, exhibiting a decay rate greater than quadratic, which is completely different from the result of slight doping. A previous work has reported that the Kondo effect negatively contributes to AHC[28, 37]. Thus, the decrease in AHC at x = 0.05, 0.08 maybe mainly because of the Kondo effect, as observed in Fe-doped Co$_{3-x}$Fe$_x$Sn$_2$S$_2$[28].

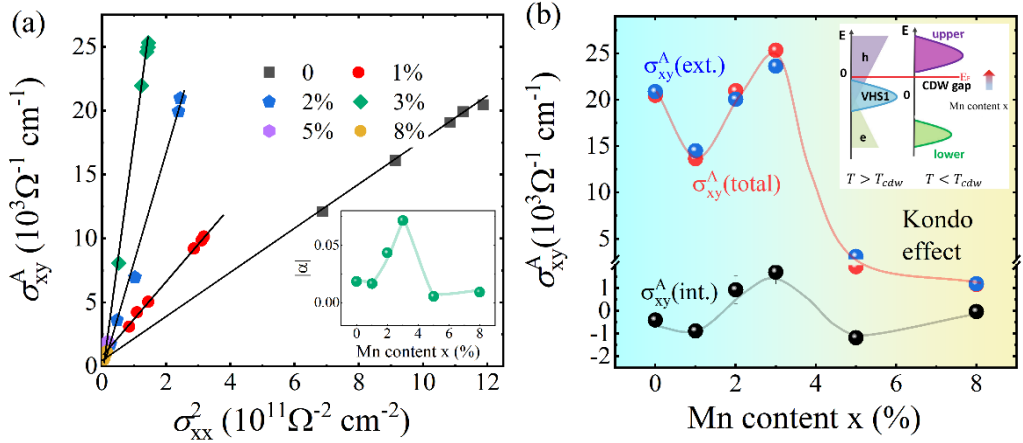

FIG. 4. (a) $\sigma_{xy}^A$ versus $\sigma_{xx}^2$ for different Mn contents. Solid lines are fittings by TYJ scaling models to extract the skew scattering constant ($\alpha$) and intrinsic AHC ($b$) for each sample. The inset shows the obtained skew scattering constant ($\alpha$). (b) Mn content dependences of intrinsic and extrinsic AHCs, separated by the TYJ model from the total measured AHC at 5 K. The inset shows the illustration of the evolution of Fermi level under different Mn contents.

In summary, a systemic study of transport properties was carried out in Mn-doped $Cs(V_{1-x}Mn_x)_3Sb_5$. As the Mn content increases, both superconductivity and CDW order are suppressed, and CDW still exists until x=0.33, which indicates an unconventional coexisting mechanism between them. In Mn-doped topological Kagome metal $CsV_3Sb_5$, significantly enhanced AHC (25331 $\Omega^{-1}cm^{-1}$) and AHA (increased by 30%) were achieved at slight doping x = 0.03. we speculate the enhanced AHC and AHA are mainly dominated by the increased asymmetric scattering of the conduction electrons from Mn, due to the effectively strengthened spin-orbital coupling. Our findings provide an effective strategy to manipulate the giant AHE by controlling the intrinsic and/or extrinsic contributions in the Kagome materials.

## Acknowledgement

This work has been supported by the National Natural Science Foundation of China (No. 12104254, No. U2430209 and No. 12425512), The Fundamental and Applied Fundamental Research Grant of Guangdong Province (Grant No. 2021B1515120015), and Guangdong Natural Science Funds for Distinguished Young Scholar (No. 2021B1515020101).